\documentclass[11pt]{article}

\usepackage{amsfonts}
\usepackage{epsfig}

\usepackage{amssymb}
\newcommand{\nn}{\nonumber}
\newcommand{\be}{\begin{equation}}
\newcommand{\bea}{\begin{eqnarray}}
\newcommand{\eea}{\end{eqnarray}}
\newcommand{\ee}{\end{equation}}

\newcommand{\eq}[1]{Eq.~(\ref{#1})}

\newcommand{\sla}[1]{\hspace{-0.1ex}\not\hspace{-0.5ex} #1\hspace{0.1ex}}

\catcode`@=11
\def\@citex[#1]#2{\if@filesw\immediate\write\@auxout{\string\citation{#2}}\fi
  \def\@citea{}\@cite{\@for\@citeb:=#2\do
    {\@citea\def\@citea{,\penalty\@m}\@ifundefined
      {b@\@citeb}{{\bf ?}\@warning
       {Citation `\@citeb' on page \thepage \space undefined}}%
\hbox{\csname b@\@citeb\endcsname}}}{#1}}

\def\citer{\@ifnextchar [{\@tempswatrue\@citexr}{\@tempswafalse\@citexr[]}}
\def\@citexr[#1]#2{\if@filesw\immediate\write\@auxout{\string\citation{#2}}\fi
  \def\@citea{}\@cite{\@for\@citeb:=#2\do
    {\@citea\def\@citea{--\penalty\@m}\@ifundefined
       {b@\@citeb}{{\bf ?}\@warning
       {Citation `\@citeb' on page \thepage \space undefined}}%
\hbox{\csname b@\@citeb\endcsname}}}{#1}}
\catcode`@=12

\begin{document}

\title{ 
\begin{flushright}
{\normalsize     
MPP-2005-147\\
{\tt hep-ph/0511248}}
\end{flushright}
\vspace*{1cm}
\bf The CP-asymmetry in resonant leptogenesis\\[8mm]}
\author{A.~Anisimov$^{\rm a}$, A.~Broncano$^{\rm b}$ and M.~Pl\"umacher$^{\rm b}$\\[1ex]
$^{\rm a}$ Institut de Th\'eorie des Ph\'enom\`enes Physiques, 
Ecole Polytechnique\\ F\'ed\'erale de Lausanne, CH-1015 Lausanne, Switzerland\\[1ex]
$^{\rm b}$ Max Planck Institute for Physics, F\"ohringer Ring 6, 80805 Munich, Germany}
\maketitle

\thispagestyle{empty}

\begin{abstract}
We study the resonantly enhanced $CP$-asymmetry in the decays
of nearly mass-degenerate heavy right-handed Majorana neutrinos
for which different formulae have been presented in the literature,
depending on the method used to calculate it. We consider two
different techniques and show that they lead to the same result,
thereby reconciling the different approaches.
\end{abstract}

\section{Introduction}
Leptogenesis \cite{FY} offers a simple and elegant explanation
for the matter-antimatter asymmetry of the universe and relates
the observed baryon asymmetry to properties of neutrinos. In
particular, a lower bound on the mass of the heavy neutrino whose
decays create the baryon asymmetry of $\sim10^9\,$GeV has been derived
in the simplest scenario of thermal leptogenesis with hierarchical
right-handed neutrinos \cite{Sacha,cmb}. Hence, the required
reheating temperature for successful leptogenesis in such a scenario
cannot be much lower \cite{analytical}, which, in supersymmetric
scenarios, may be in conflict with upper bounds on the reheating
temperature from the gravitino problem \cite{gravitino}.

Resonant leptogenesis \citer{FPSW,Pil1} has been
proposed as a way to evade this bound. 
If the heavy
right-handed neutrinos are nearly degenerate in mass, self-energy
contributions to the $CP$-asymmetries in their decays may be
resonantly enhanced. This in turn would make thermal leptogenesis
viable at much lower temperatures in the early universe.
Self-energy contributions to the $CP$-asymmetry in leptogenesis
have
been
considered numerous times in the past
\citer{FPSW,PU}. 
However, different formulae for the $CP$-asymmetry can be found in
the literature depending on the methods and approximations used to
derive it. Indeed, the correct treatment of self-energy contributions
for a decaying particle is not obvious.

$CP$-violation in the decays of heavy neutrinos arises due to
the interference of the imaginary phases of the couplings with the
absorptive parts of one-loop diagrams. A popular technique
to calculate the contribution of self-energy diagrams to
the $CP$-asymmetry 
is the use of an effective Hamiltonian, similar to that applied in 
the $K_0-\overline{K}_0$ system
\cite{FPSW,CR97,CRV96,FPS,FP}. Due to the unstable nature of
the heavy neutrinos, this approach suffers from several
shortcomings \cite{Pil0,RM}.
It is well-known that unstable particles cannot be described as
asymptotic free states, i.e.\ they cannot appear as in- or
out-states of S-matrix elements \cite{Veltman}.
Refs.~\cite{Pil1,BP,PU} tackle the problem using a
field-theoretical approach where the $CP$-asymmetry is extracted
from the resonant contributions of heavy neutrinos to stable
particle scattering amplitudes
. There, the unstable nature of
the right-handed neutrinos is taken into account by the resummation
of self-energy diagrams.
Starting from the same resummed propagator, the authors
of Refs.~\cite{Pil1,PU} and those of Ref.~\cite{BP} obtain different
final expressions for the $CP$-asymmetries in the decays
of heavy neutrinos. The main difference between these papers is
the way that the contributions from different heavy neutrinos are
inferred from the scattering amplitudes of stable particles.
In Refs.~\cite{Pil1,PU}, the contributions from different
neutrino mass eigenstates are identified by means of an
expansion of the resummed propagator around its poles, whereas
in Ref.~\cite{BP} the resummed propagator is diagonalized in
order to identify one-loop contributions to the decay amplitudes
of the heavy neutrinos.

In this paper, we compare the diagonalization and the pole expansion
methods and show that, using the same renormalization scheme, they
lead to the same result for the $CP$-asymmetry, consistent with the
one obtained in Ref.~\cite{BP}. We also discuss the range of validity
of this perturbative resummation approach. In section \ref{sec_self}
we start by introducing some notation and discuss the resummed
heavy neutrino propagator. In section \ref{scatterings} we then
introduce scattering amplitudes of stable particles from which
properties of the unstable right-handed neutrinos can be extracted
and compute the $CP$-asymmetries in their decays both in the pole
expansion and the propagator diagonalization methods,
showing that both approaches yield the same results for physical
quantities, e.g.\ decay widths and $CP$-asymmetries.

\section{Self-energy corrections to the heavy neutrino propagator}
\label{sec_self}
Leptogenesis is based on the type I seesaw model \cite{seesaw}, which supplements
the standard model with $n'$ right-handed neutrinos. The corresponding
Yukawa couplings and masses of charged leptons and neutrinos are then
given by the following Lagrangian:
\bea
    {\cal L}_Y = \overline{l_L}\,\phi\,h^*_l\,e_R
            +\overline{l_L}\,\widetilde{\phi}\,h^*_{\nu}\,N_R
            -{1\over2}\,\overline{N^C_R}\,M\,N_R
            +\mbox{ h.c.}\;.
    \label{Lagrange}
\eea
The matrices $h_{l}$, $h_{\nu}$ and $M$ are, respectively, $3 \times
3$, $3 \times n'$ and $n' \times n'$ complex matrices.  Without loss
of generality, one can always choose a basis where $h_{l}$ and $M$ are
diagonal with real and positive eigenvalues, whereas $h_{\nu}$ depends
on $3+n'+3n'$ real quantities and $3(n'-1)$ imaginary phases
\cite{Branco,BGJ}. The physical mass eigenstates are then the Majorana
neutrinos $N_i={N_{R}}_i + {N^C_R}_i$ with mass eigenvalues $M_i$. At
tree level their inverse propagator matrix reads
\be 
D_{ii}(p)\equiv S^{-1}_{ii}(p)=\sla{p}-M_i\;, 
\label{S0} 
\ee
which has poles at $p^2=M_i^2$, corresponding to stable particles.
The finite lifetime of the physical Majorana neutrinos is taken
into account by resumming self-energy diagrams.
At one loop, these diagrams introduce flavour non-diagonal
elements in the inverse propagator
\bea 
D_{ij}(p)\equiv \left(S^{-1}(p)\right)_{ij}
= {\sla{p}-M_i-\Sigma_{ij}(p)}\;, \label{S}
\eea 
where 
\be 
\label{self}
 \Sigma_{ij}(p)=
\sla{p}\,P_R\,\Sigma^R_{ij}(p^2)
      +\sla{p}\,P_L\,\Sigma^{L}_{ij}(p^2)\;,
       \label{sigma2}
\ee 
are the bare self-energies and $P_{R,L}=\frac{1\pm\gamma_5}{2}$ are
the usual chiral projectors. 

The self-energies can be written in terms
of a complex function $a(p^2)$ and a hermitian matrix $K$, 
\be
   \Sigma^L_{ij}(p^2)=\Sigma^{R}_{ji}(p^2)=a(p^2)\,K_{ij}\; ,\quad
     K_{ij}\equiv(h_{\nu}^{\dag}h_{\nu})_{ij}\; .\label{def_aK}
\ee 
In dimensional regularization, with $n=4-2\epsilon$ dimensions,
$a(p^2)$ is given by
\bea 
\label{aq2} a(p^2) = {1\over 16\pi^2}
\left(-\Delta+\ln{\left(|p^2|\over \mu^2\right)} - 2 - i\,\pi\,\theta(p^2)\right)\;,
\eea
where the ultraviolet divergence is contained in
\begin{equation}
  \Delta={1\over\epsilon}-\gamma_E+{\rm ln}(4\pi)\;.
\end{equation}

In order to identify the physical states,
the one-loop
resummed propagator has to be renormalized.
We will use the on-shell (OS) scheme, since in it the particle masses
are renormalized so as to represent the physical masses at the poles
of the propagators. We will follow the formalism for mixing renormalization
worked out in Ref.~\cite{Bernd} and the detailed computation of the renormalization
in our case can be found in Appendix A.

The renormalized inverse propagator is then given by 
\begin{eqnarray} 
\hat D(p)&=&\hat S^{-1}(p)
 \\[1ex]
&=&  \sla{p}\,P_R\,\left(1-\hat\Sigma^R(p^2)\right)
   + \sla{p}\,P_L\,\left(1-(\hat\Sigma^R(p^2))^T\right)
   - P_R\,\left( \hat M+\hat\Sigma^{M}\right)
   - P_L\,\left( \hat M+\hat\Sigma^{M^*}\right)\nonumber\;.
\label{ren_S} 
\end{eqnarray}
The off-diagonal ($i\neq j$) renormalized self-energies read
\begin{eqnarray}
\hat\Sigma_{ij}^R(p^2)&=& \frac{K_{ji}}{16\pi^2}
\left[\ln\left(\frac{|p^2|}{\hat M_i\hat M_j}\right) 
-\frac{1}{2}\frac{\hat M_j^2+\hat M_i^2}{\hat M_j^2-\hat M_i^2}\,
\ln\left(\frac{\hat M_j^2}{\hat M_i^2}\right)- i\,\pi\,\theta(p^2)
\right]\label{Sigmaij}\\
&&-\frac{K_{ij}}{16\pi^2}\,
\frac{\hat M_j\,\hat M_i}{\hat M_j^2-\hat M_i^2}\,
\ln\left(\frac{\hat M_j^2}{\hat M_i^2}\right)\;,
\nonumber\\[1ex]
\hat\Sigma_{ij}^M&=& \frac{1}{16\pi^2}\,
\frac{\hat M_j\,\hat M_i}{\hat M_j^2-\hat M_i^2}\,
\ln\left(\frac{\hat M_j^2}{\hat M_i^2}\right)\,
\left[\hat M_i\,K_{ij}+\hat M_j\,K_{ji}\right]\;,\label{SigmaMij}
\end{eqnarray}
whereas the flavour-diagonal ones are given by
\begin{eqnarray}
\hat\Sigma_{ii}^R(p^2)&=& \frac{K_{ii}}{16\pi^2}
\left[\ln\left(\frac{|p^2|}{\hat M_i^2}\right) 
- 2-i\,\pi\,\theta(p^2)\right]\;,
\label{Sigmaii}\\[1ex]
\hat\Sigma_{ii}^M&=& \frac{\hat M_i\,K_{ii}}{8\pi^2}\;.\label{SigmaMii}
\end{eqnarray}

In order to compute the elements of the renormalized propagator $\hat S_{ij}$
it is useful to decompose $\hat S(p)$ into its four chiral components,
\bea
\label{decomp}
&& \hat S(p) = P_R\,\hat S^{RR}(p^2)+P_L\,\hat S^{LL}(p^2)
           + P_L\,\sla{p}\,\hat S^{LR}(p^2)
           + P_R\,\sla{p}\,\hat S^{RL}(p^2)\;.
\eea
These chiral parts of the propagator are obtained by inserting this
decomposition into the identity
\bea 
 \hat D(p)\,\hat S(p)\;={\mathbb I}\;, 
\label{trick1}
\eea 
and multiplying from the left and the right with chiral projectors $P_{L,R}$.
The solution reads
\bea
\label{chiral_RR}
\hat S^{RR}(p^2)&=&\left[
\left(1-\hat\Sigma^{L}(p^2)\right)
\frac{p^2}{\hat M+\hat \Sigma^{M\,*}}
\left(1-\hat\Sigma^{R}(p^2)\right)-
\left(\hat M+\hat \Sigma^{M}\right)\right]^{-1}\;,\\[1ex]
\label{chiral_LL}
\hat S^{LL}(p^2)&=&\left[
\left(1-\hat\Sigma^{R}(p^2)\right)
\frac{p^2}{\hat M+\hat \Sigma^{M}}
\left(1-\hat\Sigma^{L}(p^2)\right)-
\left(\hat M+\hat \Sigma^{M\,*}\right)\right]^{-1}\;,
\\[1ex]
\label{chiral_RL}
\hat S^{RL}(p^2)&=&\frac{1}{\hat M+\hat \Sigma^{M}}
\left(1-\hat\Sigma^{L}(p^2)\right)\,\hat S^{LL}(p^2)\;,
\\[1ex]
\label{chiral_LR}
\hat S^{LR}(p^2)&=&\frac{1}{\hat M+\hat \Sigma^{M\,*}}
\left(1-\hat\Sigma^{R}(p^2)\right)\,\hat S^{RR}(p^2)\;.
\eea

For simplicity, we will restrict ourselves, in the following, 
to the case of two right-handed
neutrinos, $n'=2$. However, the generalization
to more than two generations is straightforward.

Only one-loop self-energy diagrams have been taken into consideration here, 
i.e.\ in a consistent computation the
chiral parts of the propagator have to be linearized in
 the couplings $K_{ij}$. 
For future
use, we introduce here an expansion parameter $\alpha$ related to
the largest of the couplings $K_{ij}$,
\be
\alpha= {\rm Max}\,\left[\frac{K_{ij}}{16\pi^2}\right]\,.
\ee
In the interesting case that the masses of the right-handed neutrinos are
quasi-degenerate, i.e.\ $\hat M_2-\hat M_1\ll \hat M_1$, one can define
an additional small expansion parameter
\be
\label{Delta}
\Delta\equiv \frac{\hat M_2-\hat M_1}{\hat M_1}\,.
\ee
Our results, to be presented in the following, 
will only be valid as long as $\Delta\gg\alpha$, since otherwise
perturbation theory breaks down.

To leading order in $K$, the $RR$ part of the propagator is then given by
\bea
\label{SRR}
\hat S^{RR}(p^2)= 
\left(
\begin{array}{cc}
\frac{(1+\hat\Sigma^R_{11})\,\sqrt{s_1}}{p^2-s_1}\,
& \frac{\left(\hat M_2 \hat \Sigma_{12}^R+\hat M_1 \hat \Sigma_{21}^R+\hat \Sigma_{12}^{M*}\right)\,p^2 
+\hat M_1\hat M_2 \hat \Sigma_{12}^M}
{(p^2-s_1)(p^2-s_2)}\vspace{0.5cm}\\
 \frac{\left(\hat M_2 \hat \Sigma_{12}^R+\hat M_1 \hat \Sigma_{21}^R+\hat \Sigma_{12}^{M*}\right)\,p^2 
+\hat M_1\hat M_2 \hat \Sigma_{12}^M}
{(p^2-s_1)(p^2-s_2)}
&
\frac{(1+\hat\Sigma^R_{22})\,\sqrt{s_2}}{p^2-s_2}
\end{array}
\right)\,,
\eea
where $s_{1,2}$ are the poles of the propagator. These poles are given
by the zeroes of the determinants of the inverse propagators, e.g.\ by
solving ${\rm det}(S^{RR})^{-1}=0$.
They are the same for all four chiral propagator elements and to leading
order in $K$ they read
\bea
\label{si}
s_i (p^2)= \hat M_i^2+2\,\hat M_i\,\hat\Sigma_{ii}^M+
2\,\hat M_i^2\,\hat\Sigma_{ii}^R(p^2)
=\hat M_i^2\,
\Bigg\{1+\frac{K_{ii}}{8\pi^2}
\left[\ln\left(\frac{|p^2|}{\hat M_i^2}\right)-
i\,\pi\,\theta(p^2)\right]\Bigg\}\
\eea
and, therefore,
\be
\sqrt{s_i}\equiv \hat M_i+\,\hat\Sigma_{ii}^M +
\,\hat M_i\,\hat\Sigma_{ii}^R(p^2)\,.
\ee
Note that on-shell, i.e.\ setting  $p^2= \hat M_i^2$, the poles have the familiar Breit-Wigner form
\bea
\label{si_os}
s_i (M_i^2)= \hat M_i^2 -i\,\hat M_i\,\Gamma_{i}\,,
\eea
where $\Gamma_i\equiv \frac{\hat M_i K_{ii}}{8\pi}$ are the decay widths of
the right-handed neutrinos.

Analogously, the other chiral parts of the propagator are evaluated to be
\bea
\label{SLL}
\hat S^{LL}(p^2)&=& 
\left(
\begin{array}{cc}
\frac{ (1+\hat\Sigma^R_{11})\,\sqrt{s_1} }
{p^2-s_1}\,
& \frac{\left(\hat M_2 \hat \Sigma_{21}^R+\hat M_1 \hat \Sigma_{12}^R+\hat \Sigma_{12}^M\right)\,p^2 
+\hat M_1\hat M_2 \hat \Sigma_{12}^{M*}}
{(p^2-s_1)(p^2-s_2)}\vspace{0.5cm}
\\
 \frac{\left(\hat M_2 \hat \Sigma_{21}^R+\hat M_1 \hat \Sigma_{12}^R+\hat \Sigma_{12}^M\right)\,p^2 
+\hat M_1\hat M_2 \hat \Sigma_{12}^{M*}}
{(p^2-s_1)(p^2-s_2)}
&
\frac{ (1+\hat\Sigma^R_{22})\,\sqrt{s_2} }
{p^2-s_2}
\end{array}
\right),
\hspace{0.3cm}
\\\nn\\\nn\\
\label{SLR}
\hat S^{LR}(p^2)&=& 
\left(
\begin{array}{cc}
\frac{1+\hat\Sigma^R_{11}}
{p^2-s_1}\,
& \frac{\hat M_1\hat M_2 \hat \Sigma_{12}^R+ p^2\,\hat\Sigma_{21}^R
+\hat M_1\hat \Sigma_{12}^{M*}+\hat M_2 \hat \Sigma_{12}^M}
{(p^2-s_1)(p^2-s_2)}\vspace{0.5cm}\\
 \frac
{\hat M_1\hat M_2 \hat \Sigma_{21}^R+ p^2\,\hat \Sigma_{12}^R
+\hat M_1\hat \Sigma_{12}^{M}+\hat M_2 \hat \Sigma_{12}^{M*}}
{(p^2-s_1)(p^2-s_2)}
&
\frac{1+\hat\Sigma^R_{22}}
{p^2-s_2}
\end{array}
\right)\,,
\eea
and $S^{RL}=(S^{LR})^T$.

 \section{Two-body scatterings}
\label{scatterings}

Since the right-handed neutrinos are unstable, they cannot be treated as
asymptotic free states, i.e.\ they cannot appear as in- or out-states of
S-matrix elements. Their properties can, however, be inferred from
transition matrix elements of scatterings of stable particles \cite{Veltman}.
Here, we will only consider one-loop self-energy contributions to
these scattering processes. Effects from one-loop
corrections to the vertices will be neglected in the following,
since their contribution to the
$CP$-asymmetry is well known \cite{CRV96,Pluemi1} and not
controversial.

For the case at hand, the resummed right-handed neutrino propagator
appears in the
following four lepton-Higgs scattering processes \cite{BP}:
\begin{itemize}
\item{\it Lepton-number conserving scatterings:} 
The process
$ l_\alpha\,\phi \to l_\beta\,\phi$ and its charge conjugate
$ l_\alpha^c\,\phi^*\to l_\beta^c\,\phi^*$ are mediated by heavy neutrinos.
The contribution of the resummed neutrino propagator to the amplitude for 
$ l_\alpha\,\phi \to l_\beta\,\phi$ can be written as
\bea
\label{ampli_RL}
 i\,{\cal M}&=&\overline{u}_\beta\,P_R\, \hat
h^*_{\beta i}\,\hat S_{ij}(p^2)\,\hat h_{\alpha j}\,P_L\,u_\alpha
=\overline{u}_\beta\,P_R\, \hat
h^*_{\beta i}\sla{p}\,\hat S^{RL}_{ij}(p^2)\,\hat h_{\alpha j}\,P_L\,u_\alpha\,,
\eea 
where $\hat h^*_{\alpha i}$ denote the renormalized Yukawa couplings
of right-handed neutrinos to light lepton and Higgs doublets\footnote{For
simplicity, we drop the subscript $\nu$ from the renormalized neutrino
Yukawa couplings. Further, Greek indices $\alpha,\beta,\ldots$ denote
the generation indices of SM lepton doublets, whereas Latin indices
$i,j,\ldots$ are flavour indices of the right-handed neutrinos.}.
Note that only the chiral part $\hat S^{RL}$ of the full propagator
contributes
to this amplitude.

Analogously, the contribution of $N_R$ to the amplitude for the process
$ l^c_\alpha\,\phi^* \to l^c_\beta\,\phi^*$ is given by
\bea
\label{ampli_LR}
 i\,{\cal M}&=&\overline{u}_\beta\,P_L\, \hat
h_{\beta i}\,\hat S_{ij}(p^2)\,\hat h^*_{\alpha j}\,P_R\,u_\alpha
=\overline{u}_\beta\,P_L\, \hat
h_{\beta i}\sla{p}\,\hat S^{LR}_{ij}(p^2)\,\hat h^*_{\alpha j}\,P_R\,u_\alpha\;,
\eea 
i.e.\ only $\hat S^{LR}$ contributes to this amplitude.

\item{ \it Lepton-number violating scatterings:} 

The $\Delta L =2$\
scatterings $ l_\alpha^c\,\phi^*\to l_\beta\,\phi$ and 
$l_\alpha\,\phi\to l_\beta^c\,\phi^*$ again result from right-handed
neutrino exchange.

The amplitude for the process $l^c_\alpha\,\phi^*\to l_\beta\,\phi$ 
reads
\bea
\label{ampli_RR}
 i\,{\cal M}&=&\overline{v}_\beta\,P_R\, \hat
h^*_{\beta i}\,\hat S_{ij}(p^2)\,\hat h^*_{\alpha j}\,P_R\,v_\alpha
=\overline{v}_\beta\,P_R\, \hat
h^*_{\beta i}\,\hat S^{RR}_{ij}(p^2)\,\hat h^*_{\alpha j}\,P_R\,v_\alpha\,.
\eea 
Similarly, the amplitude for the $CP$-conjugated process 
$l_\alpha\,\phi\to l_\beta^c\,\phi^*$ is
\bea
\label{ampli_LL}
 i\,{\cal M}&=&\overline{v}_\beta\,P_L\, \hat
h_{\beta i}\,\hat S_{ij}(p^2)\,\hat h_{\alpha j}\,P_L\,v_\alpha
=\overline{v}_\beta\,P_L\, \hat
h_{\beta i}\,\hat S^{LL}_{ij}(p^2)\,\hat h_{\alpha j}\,P_L\,v_\alpha\,,
\eea 
i.e.\ only $\hat S^{RR}$ and $\hat S^{LL}$ contribute to these amplitudes.
\end{itemize}

Hence, each of the chiral parts of the propagator 
participates in a
different scattering process. In the following, we will analyze the
contributions of heavy neutrinos to these scattering processes and
attempt to identify those of each mass eigenstate. This
will allow us to define effective couplings of the heavy neutrinos to
light lepton and Higgs doublets and, therefore, lead to a consistent
computation of the self-energy contribution to the $CP$-asymmetry in
heavy neutrino decays.

Different techniques of identifying the contributions of each 
neutrino mass eigenstate have been advocated in the literature.
In Ref.~\cite{PU} a decomposition into partial fractions in $p^2-s_i$
was proposed and the contributions of each mass eigenstate were
identified with those associated with the corresponding poles.
Alternatively, one can diagonalize the different chiral parts
of the propagator and identify the eigenstates of the propagator
with the effective couplings of the right-handed neutrinos, as
proposed in Ref.~\cite{BP}. In the following we will consider
both methods and will show that they lead to consistent results.

\subsection{Propagator pole expansion}
\label{polex}
Each chiral part of the resummed propagator can be decomposed into
partial fractions,
\be 
\label{XY} 
S^{AB}=\frac{X^{AB}}{p^2-s_1}+ \frac{Y^{AB}}{p^2-s_2}\,, \qquad 
\mbox{where }A,B= R,L\;,
\ee
$s_1$ and $s_2$ are the poles and $X$ and $Y$ are the matrices
which contain the coefficients of the expansion.  Note, that they are
matrices in flavour space and have no spinorial structure. The diagonal
elements of $X$ and $Y$ can be abbreviated by
\begin{eqnarray}
x_{11}&=&1+\frac{1}{2}\hat \Sigma^R_{11}(p^2)\;,\\[1ex]
y_{22}&=&1+\frac{1}{2}\hat \Sigma^R_{22}(p^2)\;,
\end{eqnarray}
whereas the non-diagonal elements are given by
\begin{eqnarray}
x_{12}&=&
-\, \frac{\hat M_1^2 \hat \Sigma_{21}^R+\hat M_1 \hat M_2 \hat \Sigma_{12}^R+
\hat M_1 \hat \Sigma_{12}^{M*}+\hat M_2 \hat \Sigma_{12}^M}
{s_2-s_1}\;,\\[1ex]
x_{21}&=&
-\, \frac{\hat M_1^2\hat \Sigma_{12}^R+\hat M_1\hat M_2\hat \Sigma_{21}^R+
\hat M_1 \hat \Sigma_{12}^{M}+\hat M_2 \hat \Sigma_{12}^{M*}}
{s_2-s_1}\;,\\[1ex]
y_{12}&=&
\
\frac{\hat M_2^2\hat \Sigma_{12}^R+\hat M_1\hat M_2\hat \Sigma_{21}^R+
\hat M_2 \hat \Sigma_{12}^{M*}+\hat M_1 \hat \Sigma_{12}^M}
{s_2-s_1}\;,\\[1ex]
y_{21}&=& \
\frac{\hat M_2^2\hat \Sigma_{21}^R+\hat M_1\hat M_2\hat \Sigma_{12}^R+
\hat M_2 \hat \Sigma_{12}^M+\hat M_1 \hat \Sigma_{12}^{M*}}
{s_2-s_1}\;.
\end{eqnarray}
With these abbreviations the coefficient matrices of the partial
fraction decomposition have a rather simple structure.  From
Eqs.~(\ref{SRR}) and (\ref{XY}) for example, one obtains for the $RR$
part of the propagator
\begin{equation}
\label{XRR}
X^{RR}=
\sqrt{s_1}\,
\left(
\begin{array}{cc}
  (x_{11})^2 & x_{12} \\[1ex]
  x_{12} & 0
\end{array}
\right)
\quad\mbox{and}\quad
Y^{RR}=
\sqrt{s_2} \,
\left(
\begin{array}{cc}
  0 & y_{12} \\[1ex]
  y_{12} & (y_{22})^2
\end{array}
\right)\;.
\end{equation}
It is clear that these results, derived in perturbation theory,
are only valid as long as the non-diagonal elements are small,
i.e.\ as long as $\Delta\gg\alpha$, as mentioned in section~\ref{sec_self}.
From  Eqs.~(\ref{SLL}) and (\ref{XY}), one analogously finds
for the $LL$ part
\begin{equation}
\label{XLL}
X^{LL}=
\sqrt{s_1}\,
\left(
\begin{array}{cc}
  (x_{11})^2 & x_{21} \\[1ex]
  x_{21} & 0
\end{array}
\right)
\quad\mbox{and}\quad
Y^{LL}=
\sqrt{s_2} \,
\left(
\begin{array}{cc}
  0 & y_{21} \\[1ex]
  y_{21} & (y_{22})^2
\end{array}
\right)\;.
\end{equation}
Further, Eqs.~(\ref{SLR}) and (\ref{XY}) yield
\begin{equation}
\label{XLR}
X^{LR}=
\left(
\begin{array}{cc}
  (x_{11})^2 & x_{12} \\[1ex]
  x_{21} & 0
\end{array}
\right)
\quad\mbox{and}\quad
Y^{LR}=
\left(
\begin{array}{cc}
  0 & y_{21} \\[1ex]
  y_{12} & (y_{22})^2
\end{array}
\right)\;.
\end{equation}
Finally, the $RL$ part of the propagator is just given by the transpose
of the $LR$ part, 
 $X^{RL}=(X^{LR})^T$ and $Y^{RL}=(Y^{LR})^T$.

The flavour structures appearing in the scattering
matrix elements (\ref{ampli_RL})-(\ref{ampli_LL}) can then be decomposed
into different contributions from each right-handed neutrino mass
eigenstate. For example, the flavour structure appearing in
Eq.~(\ref{ampli_LL}) can be written as
\begin{equation}
\hat h_{\beta i}\,\hat S^{LL}_{ij}(p^2)\,\hat h_{\alpha j}\equiv
\frac{\sqrt{s_1}}{p^2-s_1}\,\lambda_{\alpha1}\,\lambda_{\beta1}+
\frac{\sqrt{s_2}}{p^2-s_2}\,\lambda_{\alpha2}\,\lambda_{\beta2}\;,
\label{ampli_LL2}
\end{equation}
where we have introduced an effective one-loop coupling $\lambda_{\alpha i}$
of the right-handed neutrino $N_i$ to the lepton doublet $l_{\alpha}^c$
and the Higgs doublet $\phi^*$,
\begin{eqnarray}
\label{eff_h_LL1} 
\lambda_{\alpha1}(p^2)&=& 
  \hat h_{\alpha 1}\,x_{11}+\hat h_{\alpha 2}\,x_{21}\\[1ex]
  &=& \hat h_{\alpha 1}\,\left(1+\frac{1}{2}\hat \Sigma^R_{11}(p^2)\right)
      -\,\hat h_{\alpha 2}\,
      \frac{\hat M_1^2\hat \Sigma_{12}^R(p^2)+\hat M_1\hat M_2\hat \Sigma_{21}^R(p^2)
      +\hat M_1 \hat \Sigma_{12}^{M}+\hat M_2 \hat \Sigma_{12}^{M*}}
      {s_2(p^2)-s_1(p^2)}\;,\nonumber\\[1ex]
\label{eff_h_LL2} 
\lambda_{\alpha 2}(p^2)&=& 
  \hat h_{\alpha 2}\,y_{22}+\hat h_{\alpha 1}\,y_{21}\\[1ex]
  &=& \hat h_{\alpha 2}\,\left(1+\frac{1}{2}\hat \Sigma^R_{22}(p^2)\right)
      +\,\hat h_{\alpha 1}\,
      \frac{\hat M_2^2\hat \Sigma_{21}^R(p^2)+\hat M_1\hat M_2\hat \Sigma_{12}^R(p^2)
      +\hat M_2 \hat \Sigma_{12}^M+\hat M_1 \hat \Sigma_{12}^{M*}}
      {s_2(p^2)-s_1(p^2)}\,.\nonumber
\end{eqnarray}
Analogously, the flavour structure in the scattering amplitude (\ref{ampli_RR})
can be decomposed as
\begin{equation}
\hat h^*_{\beta i}\,\hat S^{RR}_{ij}(p^2)\,\hat h^*_{\alpha j}\equiv
\frac{\sqrt{s_1}}{p^2-s_1}\,\overline{\lambda}_{\alpha1}\,\overline{\lambda}_{\beta1}+
\frac{\sqrt{s_2}}{p^2-s_2}\,\overline{\lambda}_{\alpha2}\,\overline{\lambda}_{\beta2}\;,
\label{ampli_RR2}
\end{equation}
where the effective one-loop couplings $\overline{\lambda}_{\alpha i}$ of
$N_i$ to the lepton doublet $l_{\alpha}$ and the Higgs doublet $\phi$ read
\bea 
\label{eff_h_RR1}
\overline{\lambda}_{\alpha 1}(p^2)&=&  
\hat h^*_{\alpha 1}\,x_{11}+\hat h^*_{\alpha 2}\,x_{12}\vspace{0.2cm}\\
&=&\hat h^*_{\alpha 1}\,
\left(1+\frac{1}{2}\hat \Sigma^R_{11}(p^2)\right)
-\,\hat h^*_{\alpha 2}\,
\frac{\hat M_1^2 \hat \Sigma_{21}^R(p^2)+\hat M_2 \hat M_1 \hat \Sigma_{12}^R(p^2)+\hat M_1 \hat \Sigma_{12}^{M*}+\hat M_2 \hat \Sigma_{12}^M}
{s_2(p^2)-s_1(p^2)}\,,\nn\\\nn\\
\label{eff_h_RR2}
\overline{\lambda}_{\alpha 2}(p^2)&=& 
\hat h^*_{\alpha 2}\,y_{22}+\hat h^*_{\alpha 1}\,y_{12}\vspace{0.2cm}\\
&=&\hat h^*_{\alpha 2}\,
\left(1+\frac{1}{2}\hat \Sigma^R_{22}(p^2)\right)
+\,\hat h^*_{\alpha 1}\,
\frac{\hat M_2^2\hat \Sigma_{12}^R(p^2)+\hat M_1\hat M_2\hat \Sigma_{21}^R(p^2)+\hat M_2 \hat \Sigma_{12}^{M*}+\hat M_1 \hat \Sigma_{12}^M}
{s_2(p^2)-s_1(p^2)}\nn\,.
\eea
Note that $\overline{\lambda}_{\alpha i}\neq {\lambda}_{\alpha i}^*$ as a consequence of $CP$-violation.

The above effective one-loop couplings were derived from the lepton-number
violating scattering amplitudes (\ref{ampli_RR}) and (\ref{ampli_LL}).
In order for the above decomposition 
to be consistent, the lepton number conserving scattering amplitudes 
(\ref{ampli_RL}) and 
(\ref{ampli_LR}) must be recovered from the effective couplings $\lambda$ 
and $\overline{\lambda}$. Indeed, it is easy
to see that the corresponding flavour structures can be written as
\begin{eqnarray}
\hat h^*_{\beta i}\,\hat S^{RL}_{ij}(p^2)\,\hat h_{\alpha j}&=&
\frac{1}{p^2-s_1}\,\overline{\lambda}_{\beta1}\,\lambda_{\alpha1}+
\frac{1}{p^2-s_2}\,\overline{\lambda}_{\beta2}\,\lambda_{\alpha2}\;,
\label{ampli_RL2}\\[1ex]
\hat h_{\beta i}\,\hat S^{LR}_{ij}(p^2)\,\hat h^*_{\alpha j}&=&
\frac{1}{p^2-s_1}\,\lambda_{\beta1}\,\overline{\lambda}_{\alpha1}+
\frac{1}{p^2-s_2}\,\lambda_{\beta2}\,\overline{\lambda}_{\alpha2}\;.
\label{ampli_LR2}
\end{eqnarray}
Hence, the effective couplings in Eqs.~(\ref{eff_h_LL1}), (\ref{eff_h_LL2}), 
(\ref{eff_h_RR1}), and (\ref{eff_h_RR2}) consistently take the one-loop
self-energy contributions to couplings of right-handed neutrinos to light
lepton and Higgs doublets into account. Correspondingly, the self-energy
contributions to heavy neutrino decay widths can be written as
\bea
&&\Gamma(N_i\to l\,\phi)= 
\frac{\hat M_i}{16\pi}\,\sum_\alpha 
{\overline{\lambda}}_{\alpha i}^{\ *}(p^2)\,\overline{\lambda}_{\alpha i}(p^2)\,,
\label{Gamma1}\\[1ex]
&&\Gamma(N_i\to l^c\,\phi^*)=
\frac{\hat M_i}{16\pi}\,\sum_\alpha 
{\lambda}_{\alpha i}^*(p^2)\,{\lambda}_{\alpha i}(p^2)\;.
\label{Gamma2}
\eea

The partial decay widths of $N_1$ are then evaluated to be
\bea
\label{Gamma1_RR}
\Gamma(N_1\to l\,\phi)&=& 
\frac{\hat M_1}{16\pi}\,\Bigg\{ 
K_{11}\,|x_{11}|^2+
\,2\,{\rm Re}\,\left(K_{21}\,x_{12}\right)\Bigg\}\;,\\[1ex]
\label{Gamma1_LL}
\Gamma(N_1\to l^c\,\phi^*)&=&
\frac{\hat M_1}{16\pi}\,\Bigg\{ 
K_{11}\,|x_{11}|^2+
\,2\,{\rm Re}\,\left(K_{12}\,x_{21}\right)\Bigg\}\;.
\eea
Analogously, the partial decay widths of $N_2$ read 
\bea
\label{Gamma2_RR}
\Gamma(N_2\to l\,\phi)&=& 
\frac{\hat M_2}{16\pi}\,\Bigg\{ 
K_{22}\,|y_{22}|^2+
\,2\,{\rm Re}\,\left(K_{12}\,y_{12}\right)\Bigg\}\,,\\
\label{Gamma2_LL}
\Gamma(N_2\to l^c\,\phi^*)&=&
\frac{\hat M_2}{16\pi}\,\Bigg\{ 
K_{22}\,|y_{22}|^2+
\,2\,{\rm Re}\,\left(K_{21}\,y_{21}\right)\Bigg\}\,.\nn
\eea

It is now straightforward to compute the self-energy contributions to
the $CP$-asymmetry $\varepsilon_i$ in the decay of $N_i$. From 
Eqs.~(\ref{Gamma1}) and (\ref{Gamma2}) one finds
\begin{equation} 
\label{asym} 
\varepsilon_i(p^2)\equiv\frac
{\Gamma(N_i\to l\,\phi)-\Gamma(N_i\to l^c\,\phi^*)}
{\Gamma(N_i\to l\,\phi)+\Gamma(N_i\to l^c\,\phi^*)}=
\frac{\sum_{\alpha}\left[\left|\overline{\lambda}_{\alpha i}(p^2)\right|^2
                         -\left|{\lambda}_{\alpha i}(p^2)\right|^2\right]} 
     {\sum_{\alpha}\left[\left|\overline{\lambda}_{\alpha i}(p^2)\right|^2
                         +\left|{\lambda}_{\alpha i}(p^2)\right|^2\right]} \;.
\end{equation}
Using Eqs.~(\ref{eff_h_LL1}), (\ref{eff_h_LL2}), (\ref{eff_h_RR1}),
and (\ref{eff_h_RR2}) for the effective couplings and going on-shell,
one finally obtains for the $CP$-asymmetries
\begin{eqnarray}
\varepsilon_1\left(\hat M_1^2\right)&=&
\frac{{\rm Im}\left(K_{12}^2\right)}{8\pi\,K_{11}}\;
\frac{\hat M_1\,\hat M_2\,\left(\hat M_2^2-\hat M_1^2\right)}
  {\left(\hat M_2^2-\hat M_1^2
  -\frac{1}{\pi}\,\hat M_2\,\Gamma_2\,
  \ln\left(\frac{\hat M_2^2}{\hat M_1^2}\right)\right)^2 
  +\left(\hat M_2\,\Gamma_2-\hat M_1\,\Gamma_1\right)^2}\;,
  \label{epsilon1}\\[1ex]
\varepsilon_2\left(\hat M_2^2\right)&=&
\frac{{\rm Im}\left(K_{12}^2\right)}{8\pi\,K_{22}}\;
\frac{\hat M_1\,\hat M_2\,\left(\hat M_2^2-\hat M_1^2\right)}
  {\left(\hat M_2^2-\hat M_1^2
  -\frac{1}{\pi}\,\hat M_1\,\Gamma_1\,
  \ln\left(\frac{\hat M_2^2}{\hat M_1^2}\right)\right)^2 
  +\left(\hat M_2\,\Gamma_2-\hat M_1\,\Gamma_1\right)^2}\;.
  \label{epsilon2}
\end{eqnarray}
The logarithmic terms in the denominators of Eqs.~(\ref{epsilon1}) and
(\ref{epsilon2}) describe, e.g.\ the running of $\hat M_2$ to the scale
$p^2=\hat M_1^2$ and vice-versa. In the resonance regime, $\Delta\ll1$,
these terms are 
${\cal O}(\alpha\Delta)$
 and, therefore, they can be neglected. We have
just included them for completeness.

The main result of this computation is that the regulator of the mass
singularity at $\hat M_1=\hat M_2$ is the difference of the masses times
the decay widths of the neutrinos, i.e.\ 
$\hat M_2\,\Gamma_2-\hat M_1\,\Gamma_1$,
in perfect analogy, e.g.\ with the
$CP$-asymmetry in the $K_0$-$\overline{K}_0$ system \cite{ChengLi}. This
result is consistent with the one obtained in Ref.~\cite{BP} but deviates
from the expression for the $CP$-asymmetry given in Ref.~\cite{PU}, where
only one of the decay widths appears as regulator. 

\subsection{Diagonalization of the propagator}
An alternative approach to determining the contributions of the different
right-handed neutrino mass eigenstates to the scattering amplitudes
(\ref{ampli_RL})-(\ref{ampli_LL}) consists in diagonalizing the various
chiral parts of the propagators and identifying the diagonal elements
with the contributions from the neutrino mass eigenstates \cite{BP}.
 
The propagators $S^{RR}$ and $S^{LL}$ are complex symmetric matrices,
which are diagonalized by complex orthogonal matrices $V$ and $U$,
respectively,
\bea
\label{sdiag}
\hat S^{RR}(p^2) &=& 
Z(p^2)\,V^T(p^2) \, \hat S^{\rm diag}(p^2)\, V(p^2)\, Z(p^2)\;,\\[1ex]
\hat  S^{LL}(p^2) &=& 
 Z(p^2)\,U^T(p^2) \, \hat S^{\rm diag}(p^2)\, U(p^2)\,Z(p^2)\;.
\label{sdiag2}
\eea 
Here, $Z$ is a diagonal normalization matrix,
\be 
\label{Z} 
Z= \left(
\begin{array}{cc}
\left[\sqrt{s_1}\left(1+\hat\Sigma_{11}^R\right)\right]^{1/2}&0\\
0&\left[\sqrt{s_2}\left(1+\hat\Sigma_{22}^R\right)\right]^{1/2}\\
\end{array}
\right)\,, 
\ee
chosen in such a way as to give the canonical normalization for the
diagonal propagator
\begin{equation}
\hat S^{\rm diag} = \left(
\begin{array}{cc}
  \frac{1}{p^2-s_1} & 0\\[1ex]
  0 & \frac{1}{p^2-s_2}
\end{array}\right)\;,
\end{equation}
where $s_i$ are the propagator poles of \eq{si}.

Let us first consider the diagonalization of $\hat S^{LL}$. By choosing
\be 
\label{U} 
U= \left(
\begin{array}{cc}
\cos\theta_U&-\sin\theta_U\\
\sin\theta_U&\cos\theta_U\\
\end{array}
\right)\,, 
\ee 
the complex mixing angle $\theta_U$ can be determined from
Eqs.~(\ref{SLL}) and (\ref{sdiag2})
\bea
\label{tan_thetaU}
\tan\left(2\theta_U\right)&\!\!\!\!=&\!\!\!\!
\frac{2}{(s_1(p^2) s_2(p^2))^{1/4}}\,
\frac{\,p^2 \,\left(\hat M_2 \hat \Sigma_{21}^R(p^2)+
\hat M_1 \hat \Sigma_{12}^R(p^2)
+\hat \Sigma_{12}^M\right)
+\hat M_1\hat M_2 \hat \Sigma_{12}^{M*}}
{s_2(p^2)-s_1(p^2)}\;.
\eea 
Note, that \eq{tan_thetaU} is of order
\bea
\tan(2\theta_U)&=&
\frac{{\cal O}(\alpha)}{{\cal O}(\Delta)+{\cal O}(\alpha)}\;.
\eea

Thus, if the condition for the applicability of perturbation theory,
$\Delta\gg\alpha$, is fulfilled, \eq{tan_thetaU} can be then expanded and the elements
of the mixing matrix $U$ read
\bea
\label{cos_thetaU}
\cos\theta_U&\simeq& 1\,,\\
\label{sin_thetaU}
\sin\theta_U&\simeq& \frac{1}{(s_1(p^2) s_2(p^2))^{1/4}}\,
\frac{\,p^2 \,\left(\hat M_2 \hat \Sigma_{21}^R(p^2)+\hat M_1 \hat \Sigma_{12}^R(p^2)
+\hat \Sigma_{12}^M\right)
+\hat M_1\hat M_2 \hat \Sigma_{12}^{M*}}
{s_2(p^2)-s_1(p^2)}\,.\,
\eea
In perfect analogy to the procedure followed in the pole expansion
method, the flavour structure in the scattering amplitude 
(\ref{ampli_LL}) can now be written as
\begin{equation}
\hat h_{\beta i}\,\hat S^{LL}_{ij}(p^2)\,\hat h_{\alpha j}=
\sum\limits_i \left(\hat h\,Z\,U^T\right)_{\beta i}\,  s_i^{-1/4}\,
  \sqrt{s_i}\,\ S_{ii}^{\rm diag}\,
  \left(\hat h\,Z\,U^T\right)_{\alpha i}\,s_i^{-1/4}\;,
\end{equation}
Hence, we can define an effective one-loop coupling $\xi_{\alpha i}$
of the right-handed neutrino $N_i$ to the lepton doublet $l^c_{\alpha}$
and the Higgs doublet $\phi^*$,
\begin{equation}
  \xi_{\alpha i}\equiv 
  \left(\hat h\,Z\,U^T\right)_{\alpha i}\,s_i^{-1/4}\;.
\end{equation}
From Eqs.~(\ref{Z}), (\ref{cos_thetaU}) and (\ref{sin_thetaU}) explicit
expressions for these effective couplings are given by
\begin{eqnarray}
\xi_{\alpha 1}(p^2)&=&\hat h_{\alpha 1}\,
\left(1+\frac{1}{2}\hat \Sigma^R_{11}(p^2)\right)
-\,\hat h_{\alpha 2}\,\frac{p^2\,\left(\hat M_2 \hat \Sigma_{21}^R(p^2)
+\hat M_1 \hat \Sigma_{12}^R(p^2)+\hat \Sigma_{12}^{M}\right)
+\hat M_1\hat M_2 \hat \Sigma_{12}^{M*}}
{\hat M_1\,\left(s_2(p^2)-s_1(p^2)\right)}
\label{eff_h_LL1_diag}\;,\nn\\\\
\xi_{\alpha 2}(p^2)&=&\hat h_{\alpha 2}\,
\left(1+\frac{1}{2}\hat \Sigma^R_{22}(p^2)\right)
+\,\hat h_{\alpha 1}\,\frac{p^2\,\left(\hat M_2 \hat \Sigma_{21}^R(p^2)
+\hat M_1 \hat \Sigma_{12}^R(p^2)+\hat \Sigma_{12}^{M}\right) 
+\hat M_1\hat M_2 \hat \Sigma_{12}^{M*}}
{\hat M_2\,\left(s_2(p^2)-s_1(p^2)\right)}
\label{eff_h_LL2_diag}\;.\nn\\
\end{eqnarray}

Similarly, $S^{RR}$ is diagonalized by the normalization matrix $Z$
given in \eq{Z} and the orthogonal matrix
\be 
\label{V} 
V= \left(
\begin{array}{cc}
\cos\theta_V&-\sin\theta_V\\
\sin\theta_V&\cos\theta_V\\
\end{array}
\right)\,, 
\ee
where the complex mixing angle $\theta_V$ is given by
\bea
\label{tan_thetaV}
\tan\left(2\theta_V\right)&\!\!\!\!=&\!\!\!\!
 \frac{2}{(s_1(p^2) s_2(p^2))^{1/4}}\,\frac{p^2\,\left(\hat M_1 \hat \Sigma_{21}^R(p^2)+
\hat M_2 \hat \Sigma_{12}^R(p^2)+
\hat \Sigma_{12}^{M*}\right)
+\hat M_1\hat M_2 \hat \Sigma_{12}^M}
{s_2(p^2)-s_1(p^2)}\;.
\eea 
Again, at leading order, one finds
\bea
\label{cos_thetaV}
\cos\theta_V&\simeq& 1\,,\\
\label{sin_thetaV}
\sin\theta_V&\simeq& 
 \frac{1}{(s_1(p^2) s_2(p^2))^{1/4}}\,\frac{p^2\,
\left(\hat M_1 \hat \Sigma_{21}^R(p^2)+
\hat M_2 \hat \Sigma_{12}^R(p^2)+
\hat \Sigma_{12}^{M*}\right)
+\hat M_1\hat M_2 \hat \Sigma_{12}^M}
{s_2(p^2)-s_1(p^2)}\;.
\eea
The flavour structure in the amplitude (\ref{ampli_RR}) can
be then written as
\begin{equation}
\hat h^*_{\beta i}\,\hat S^{RR}_{ij}(p^2)\,\hat h^*_{\alpha j}=
\sum\limits_i \left(\hat h^*\,Z\,V^T\right)_{\beta i}\,  s_i^{-1/4}\,
  \sqrt{s_i}\,\ S_{ii}^{\rm diag}\,
  \left(\hat h^*\,Z\,V^T\right)_{\alpha i}\,s_i^{-1/4}\;.
\end{equation}
We can again define an effective one-loop coupling
$\overline{\xi}_{\alpha i}$ of $N_i$ to the lepton doublet
$l_{\alpha}$ and the Higgs doublet $\phi$,
\begin{equation}
\overline{\xi}_{\alpha i}\equiv 
  \left(\hat h^*\,Z\,V^T\right)_{\alpha i} 
  \,s_i^{-1/4}\;.
\end{equation}
Explicitly, these effective couplings read
\begin{eqnarray}
\overline{\xi}_{\alpha 1}(p^2)&=&\hat h^*_{\alpha 1}\,
\left(1+\frac{1}{2}\hat \Sigma^R_{11}(p^2)\right)
-\,\hat h^*_{\alpha 2}\,\frac{p^2\,\left(\hat M_1 \hat \Sigma_{21}^R(p^2)+
\hat M_2 \hat \Sigma_{12}^R(p^2)
+\hat \Sigma_{12}^{M*}\right)
+\hat M_1\hat M_2 \hat \Sigma_{12}^M}
{\hat M_1\,\left(s_2(p^2)-s_1(p^2)\right)}
\;,
\label{eff_h_RR1_diag}\nn\\\\
\overline{\xi}_{\alpha 2}(p^2)&=&\hat h^*_{\alpha 2}\,
\left(1+\frac{1}{2}\hat \Sigma^R_{22}(p^2)\right)
+\,\hat h^*_{\alpha 1}\,\frac{p^2\,
\left(\hat M_1 \hat \Sigma_{21}^R(p^2)+
\hat M_2 \hat \Sigma_{12}^R(p^2)
+\hat \Sigma_{12}^{M*}\right) 
+\hat M_1\hat M_2 \hat \Sigma_{12}^M}
{\hat M_2\,\left(s_2(p^2)-s_1(p^2)\right)}
\label{eff_h_RR2_diag}\;.\nn\\
\end{eqnarray}

Note that on-shell, i.e.\ for $p^2=\hat M_i^2$. 
the effective couplings in
Eqs.~(\ref{eff_h_LL1_diag}), (\ref{eff_h_LL2_diag}), 
(\ref{eff_h_RR1_diag}) and (\ref{eff_h_RR2_diag}), agree with the
effective couplings (\ref{eff_h_LL1}), (\ref{eff_h_LL2}), 
(\ref{eff_h_RR1}) and (\ref{eff_h_RR2}), derived in the pole expansion
method:
\begin{eqnarray}
  \xi_{\alpha i}(\hat M_i^2)=\lambda_{\alpha i}(\hat M_i^2)\;,\\[1ex]
  \overline{\xi}_{\alpha i}(\hat M_i^2)=
  \overline{\lambda}_{\alpha i}(\hat M_i^2)\;.
\end{eqnarray}
Thus, the effective resummed one-loop couplings of the different
 right-handed neutrino mass eigenstates derived in the pole expansion
 and the diagonalization methods are identical on-shell.
In particular, this means that physical quantities computed on-shell,
e.g.\ decay widths and $CP$-asymmetries, 
in the propagator diagonalization
method will agree with those obtained in the pole expansion method,
thereby confirming our results from section~\ref{polex}.

\section{Conclusions}
In this paper we have studied the resonantly enhanced $CP$-asymmetry
in the decays of nearly mass-degenerate heavy right-handed neutrinos,
a regime known as resonant leptogenesis. Such a scenario is phenomenologically
interesting since it allows to evade the rather stringent lower
limit on the reheating temperature that can be obtained in the
simplest scenario of thermal leptogenesis with hierarchical
right-handed neutrino masses. Further, it may open the possibility
of directly observing right-handed neutrinos at future colliders
\cite{PilU2}.

Unfortunately, different formulae for the $CP$-asymmetry in
resonant leptogenesis had previously been proposed in the
literature. Obviously, this hampers phenomenological investigations
of resonant leptogenesis, since it was not clear which of these formulae 
one should use.

We 
have clarified 
the situation by computing the $CP$-asymmetry
with two independent methods, the pole expansion and the propagator
diagonalization method. We showed that, within the same
renormalization scheme, both methods give identical results for
physical quantities, such as decay widths and, in particular,
$CP$-asymmetries in the decays of heavy right-handed neutrinos.
Furthermore, we have also discussed the range of validity of the 
resulting formulae. Due to the limitations of the perturbative approach, 
the degree of degeneracy of heavy neutrinos must be restricted to be much
larger than the expansion parameter, determined by the neutrino Yukawa 
couplings.

\section*{Acknowledgments}

We would like to thank P.~Di Bari, S.~Dittmaier and B.~Gavela for 
for useful discussions and comments on the manuscript. A.~B.~and M.~P.~
acknowledge support by the Deutsche Forschungsgemeinschaft within
the Emmy-Noether program. 

\appendix
\section{On-shell renormalization with particle mixing}

In this appendix we will briefly summarize the renormalization procedure
we use, following the formalism developed in Ref.~\cite{Bernd}.
In the seesaw model described by the Lagrangian in \eq{Lagrange}, the bare
one-loop self-energy of the right-handed neutrinos has the structure given
in Eq.~(\ref{self}).

The renormalized masses and fields, denoted by a hat in the following, are
related to the bare ones by the counterterms:
\bea
&&\hat M_{i}= M_{i}+\delta M_{i}\;,\\[1ex]
&&\hat N_{R\,i} = \left(Z^{R}_{ij}\right)^{1/2} \,N_{R\,j} 
                = \left(\delta_{ij}+\frac{1}{2}\delta Z^R_{ij}\right)
\,N_{R\,j} 
\;,\\[1ex]
&&\hat N_{R\,i}^c= \left(Z^{R\,*}_{ij}\right)^{1/2} \,N_{R\,j}^c
                 = \left(\delta_{ij}+\frac{1}{2}\delta Z^{R\,*}_{ij}\right)\,N_{R\,j}^c\;.
\eea
From the counterterm Lagrangian, one then obtains the following relations
between renormalized and bare self-energies:
\bea
&&\hat\Sigma_{ij}^R(p^2)={\Sigma_{ij}^R}(p^2)+\frac{1}{2}\left(\delta
Z_{ij}^R+\delta Z_{ji}^{R*}\right)
\label{SigmaR}\;,\\[1ex]
&&\hat\Sigma_{ij}^M(p^2)=-\frac{1}{2}\left({\hat M_j}\,\delta
Z_{ji}^R+ {\hat M_i}\,\delta Z_{ij}^{R}\right)-\delta_{ij}\,\delta M_i\,. 
\label{SigmaM}
\eea
In the OS scheme, the counterterms are determined from the following
renormalization conditions:
\bea
\label{cond1}
\hat \Sigma^{dis}_{ij}(p)u_j(p)|_{p^2= \hat M_j^2}&=&0\;,\\[1ex]
\frac{1}{\sla{p}-\hat M_i} \hat \Sigma^{dis}_{ii}(p)u_i(p)|_{p^2\to
\hat M_i^2}&=& 0\;,
\label{cond2} 
\eea 
where the subscript $dis$ refers to the dispersive part of the self-energy,
since absorptive parts cannot contribute to the renormalization
without spoiling the required hermiticity of the counterterm Lagrangian.
The first renormalization condition, Eq.~(\ref{cond1}), yields the
following two equations:
\bea
\label{equal} 
&&\hat\Sigma_{ij}^{R\,dis}(\hat M_j^2)\,\hat M_j+
  \hat\Sigma_{ij}^{M\,dis\,*}(\hat M_j^2)=0 \;,\\
&&\hat\Sigma_{ji}^{R\,dis}(\hat M_j^2)\,\hat M_j+
  \hat\Sigma_{ij}^{M\,dis}(\hat M_j^2)=0\,.
\eea
Since $\Sigma_{ij}^{R\,dis}=\Sigma_{ji}^{R\,dis^*}$ these two equations are
equivalent. From Eqs.~(\ref{SigmaR}) and (\ref{SigmaM}) one then obtains
the mass counterterms as well as the non-diagonal elements of $\delta Z^R$,
\begin{eqnarray} 
\delta M_i &=& \hat M_i\,\Sigma^{R\,dis}_{ii}(\hat M_i^2)\;,\\[1ex]
\delta Z^R_{ij}&=& \frac{2}{\hat M^2_i-\hat M^2_j}
\left[\hat M^2_j\,\Sigma^{R\,dis}_{ij}(\hat M^2_j)+
 \hat M_j\,\hat M_i\,\Sigma^{R\,dis}_{ji}(\hat M^2_j)\right]\;,
\quad\mbox{for }i\neq j\;.
\end{eqnarray}
Similarly, Eq.~(\ref{cond2}) yields
\bea 
\label{diag}
\hat\Sigma_{ii}^{R\,dis}(\hat M_i^2)+ 2\,\hat M_i\,
\frac{\partial}{\partial
p^2}\left(\hat M_i\,\hat\Sigma_{ii}^{R\,dis}(p^2)+
    \hat\Sigma_{ii}^{M\,dis}(p^2)\right)
\Big|_{p^2=\hat M_i^2}=0\;,
\eea 
from which the flavour diagonal counterterms $\delta Z_{ii}$ can be obtained,
\bea 
\label{ZdiaR} 
\delta Z^R_{ii}&=&-\Sigma^{R\,dis}_{ii}(\hat M^2_i)-
2\,\hat M_i^2\,\frac{\partial}{\partial p^2}
\left[\Sigma^{R\,dis}_{ii}(p^2)
\right]\Big|_{p^2=\hat M_i^2}\;.
\eea
Substituting these counterterms into Eqs.~(\ref{SigmaR}) and (\ref{SigmaM}) then
gives rise to the renormalized self-energies in 
Eqs.~(\ref{Sigmaij})-(\ref{SigmaMii}).


\end{document}